\documentclass[aps,pra,twocolumn,superscriptaddress,showpacs]{revtex4}

\usepackage[english]{babel}
\usepackage{graphicx}
\usepackage{amsmath}
\usepackage{amssymb}
\usepackage{color}

\usepackage{cancel}

\newcommand{\Di}{\mathcal{D}}
\newcommand{\Ai}{\mathcal{A}}
\newcommand{\eS}{\mathcal{S}}
\newcommand{\Cor}{\mathfrak{C}}
\newcommand{\Dsla}{\cancel{\mathcal{D}}}
\newcommand{\Asla}{\cancel{\mathcal{A}}}
\newcommand{\sigsla}{\cancel{\sigma}}
\newcommand{\gspa}{\mathfrak{S}}
\newcommand{\tenstype}[2]{\begin{tiny}$\left( \!\! \begin{array}{c} #1 \\ #2 \end{array} \!\! \right)$ \end{tiny}}

\begin{document}

\title{Homogeneous binary tree as ground states of quantum critical Hamiltonians}

\author{P. Silvi}
\affiliation{International School for Advanced Studies (SISSA),
  		Via Bonomea 265, I-34136 Trieste, Italy}
\author{V. Giovannetti}
\affiliation {NEST,  Scuola Normale Superiore and Istituto di Nanoscienze - CNR,  Pisa, Italy}
\author{S. Montangero}
\affiliation {Institut f\"ur Quanteninformationsverarbeitung,
  		Universit\"at Ulm, D-89069 Ulm, Germany}
  \author{M. Rizzi}
\affiliation{Max-Planck-Institut f\"{u}r Quantenoptik, Hans-Kopfermann-Strasse 1, 85748 Garching, Germany }
\author{J. I. Cirac}
\affiliation{Max-Planck-Institut f\"{u}r Quantenoptik, Hans-Kopfermann-Strasse 1, 85748 Garching, Germany }
\author{R. Fazio}
\affiliation {NEST,  Scuola Normale Superiore and Istituto di Nanoscienze - CNR,  Pisa, Italy}
\affiliation{Center for Quantum Technologies, National University of Singapore, Republic of Singapore}

\date{\today}

\begin{abstract}
Many-body states whose wave-function admits a  representation in terms of a uniform binary-tree tensor decomposition 
are shown to obey to power-law two-body correlations functions. Any such state can be associated with the ground 
state of a translational invariant Hamiltonian  which, depending on the dimension of the systems sites, involve at most 
couplings between third-neighboring sites. Under  general conditions it is shown that 
  they describe unfrustrated systems which admit an exponentially large degeneracy of the ground state.

\end{abstract}

\pacs{03.67.-a,05.30.-d,89.70.-a}

\maketitle
\section{Introduction}\label{sec:intro} 
The selection of suitable tailored variational wavefunctions is a fundamental problem in the study of 
quantum many-body systems~\cite{CV}.  The variational ansatz
must satisfy two basic requirements: it should provide an accurate approximation of the target state (e.g. the ground state), and
 it should allow for an efficient evaluation of the relevant physical quantities (e.g. local observables and associated 
correlation functions).  Matrix Product States (MPS) are a successful example of this kind~\cite{FNW}.  It is possible to quantify their 
accuracy to approximate the exact wave-function~\cite{mpsverstra} and in some specific cases~\cite{affleck} the ground state itself 
is in a matrix product  form (e.g. see Ref.~\cite{ADV} for a review). MPS are specifically suited to deal with not 
critical, short-range, one-dimensional  Hamiltonians. In order to overcome these limitations several generalizations have been 
proposed~\cite{peps,wgs,mera,wen}. Projected entangled pair states~\cite{peps} were introduced to deal with  higher dimensions,  weighted graph states~\cite{wgs} to treat systems with long-range interactions, and  Multi-scale Entanglement Renormalization 
Ansatz (MERA)~\cite{mera}   to efficiently address critical systems.

In this work we focus on one-dimensional quantum critical systems using \emph{homogeneous} Binary-Tree States (HBTSs) as 
variational  states.  They  share some structural properties of scale-invariant MERA states  (including the possibility of constructing efficient 
optimizing algorithms~\cite{NS,VIDAL1,mera}) but admit  a simpler description  and provide a  prototypical realization of a real-space renormalization process.  Even though on general grounds it can be argued that these states  are suitable candidates to 
approximate critical systems (e.g. they violate area law~\cite{REV} with logarithmic corrections~\cite{BTvidal}) an explicit derivation of their 
critical properties  is still missing.  We will prove that HBTSs can describe critical systems  by computing the correlation functions and 
show that they decay in a power-law fashion.  

Once ascertained that HBTSs do describe critical ground state it would be important to know if there are cases in which they are 
actually the exact ground state of a given model Hamiltonian.  Despite the large body of work devoted so far on the subject
there is no definite answer for critical systems (up to now only  approximated, numerical evidences have been gathered on this 
issue).  Given their ubiquitous presence of in condensed matter and statistical mechanics, this question is of particular relevance 
both conceptually  and for possible numerical implementations.  In this work we show that  HBTSs can be associated, in the 
thermodynamic limit, to (non trivial)  local and translationally-invariant  parent Hamiltonians~\cite{newnote2}.  
Furthermore, similarly to what was done for MPS~\cite{mpskarimi,FNW},  we discuss sufficient conditions under which such 
operators continue to be  parent Hamiltonians also for finite sites. By construction this allows us to identify a class of (non trivial)  
unfrustrated Hamiltonians whose characterization has been attracting some interest recently, e.g. see Ref.~\cite{SHOR}.  
Although we concentrate only on binary trees the method  we present to construct the parent Hamiltonian can be apply to other 
tensor structures, as the MERA, which support scale invariance.

\begin{figure}
\includegraphics[width=240pt]{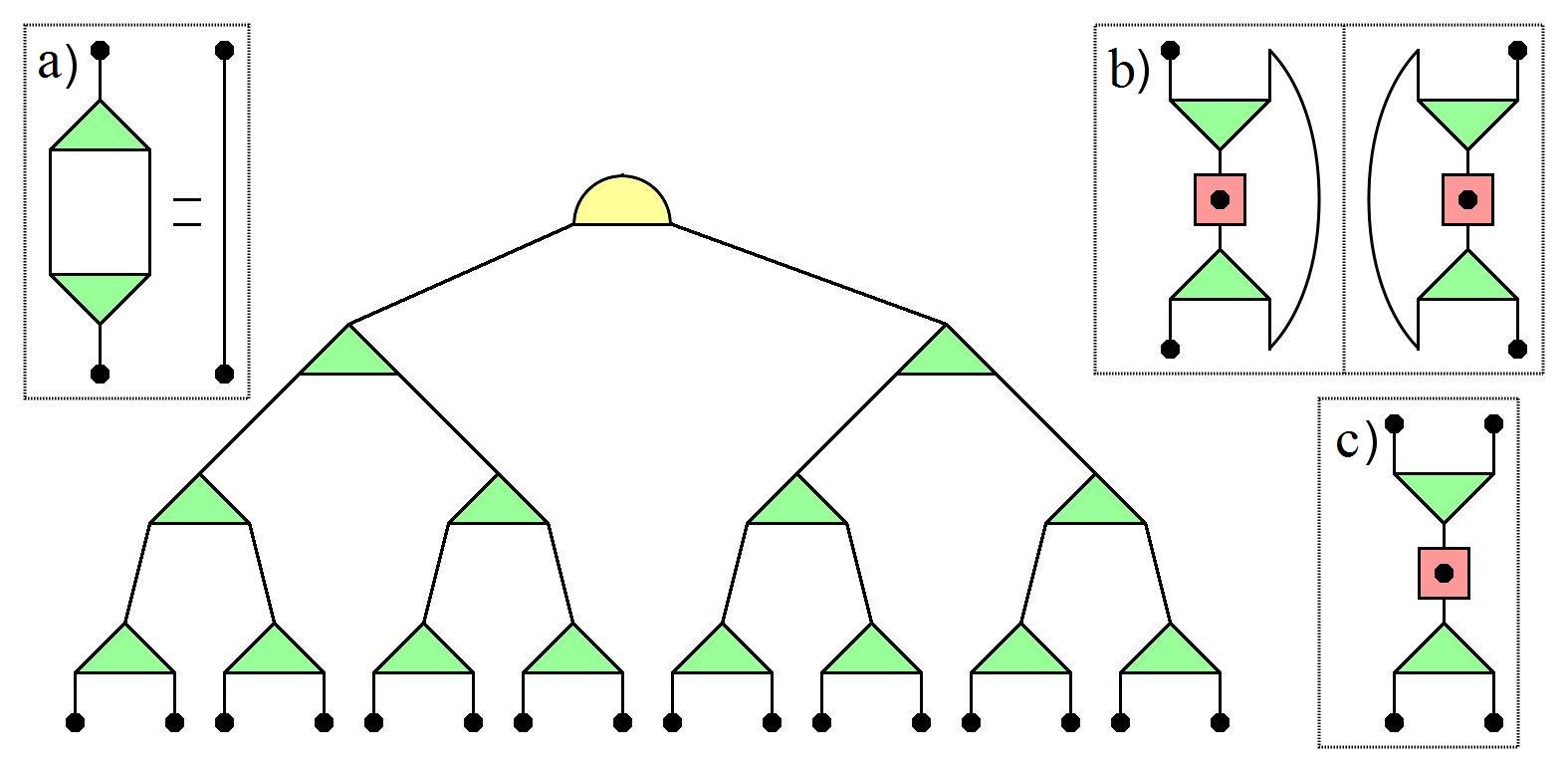} 
\caption{(Color online) BT network for $16= 2^{n}$ 
sites.
Inset A) shows  the isometric property of $\lambda$; 
B) the maps $\Di_L$ (left) and $\Di_R$;  and C) the
 map ${\cal S}$ of Eq.~\eqref{eq:fff}.}
\label{fig:alberone}
\end{figure}

The paper is organized as follows:
Sec.~\ref{sec:homo} is devoted to introduce the basic properties of HTBSs; Sec.~\ref{sec:corr} focuses on how correlation functions can be computed for such systems and shows
the critical characters of such quantities; Sec.~\ref{sec:par} discuss how to construct parent Hamiltonians for HTBSs.
Conclusions and remarks
are presented in Sec.~\ref{con}. In the appendix we discuss how to construct the parent Hamiltonian in the case of a MERA state.

\section{Homogeneous Binary Tree States}\label{sec:homo}  
Consider a 1D lattice of $N = 2^n$ sites,
of a given local dimension $d$, with periodic boundary conditions.
A generic pure state  of such system can always be expressed as 
\begin{eqnarray}
|\psi^{(n)}\rangle = \sum_{\ell_1, \ldots , \ell_N = 1}^{d} \mathcal{T}_{\ell_1, \ldots \ell_N}
 |\xi_{\ell_1} \ldots \xi_{\ell_N} \rangle\;,
 \end{eqnarray} 
with $\{ |\xi_{i}\rangle \}_i$ a canonical basis for the single qudit and
$\mathcal{T}$ a  type-\tenstype{0}{N} tensor.
HBTSs of depth $n$ are identified as those  $|\psi^{(n)}\rangle$ whose $\mathcal{T}$ can be decomposed 
in terms of  smaller tensors 
as in Fig.~\ref{fig:alberone}.  Following Ref.~\cite{mera},
each node of such graph represents a tensor (the  emerging legs of the node being its indices),
while a link connecting any two nodes represents contraction of the corresponding indices.  
The (yellow) element on the top of Fig.~\ref{fig:alberone} describes 
a type-\tenstype{0}{2} tensor ${\cal C}$
  of elements ${\cal C}_{\ell_1,\ell_2}$, 
while the $2N-1$ triangles represent the same
$d \times d^2$  tensor $\lambda$ of type-\tenstype{1}{2} whose elements ${\lambda}^{u}_{\ell_1,\ell_2}$,
satisfy the isometric condition
\begin{equation} \label{eq:isometry}
 \sum_{k_1, k_2} \lambda^{u}_{k_1, k_2} {\bar{\lambda}}_{\ell}^{k_1, k_2} = \delta^{u}_{\ell},
\end{equation} 
where $\delta^{u}_{\ell}$ is the Kronecker delta and  $\bar{\lambda}_{\ell}^{u_1,u_2} \equiv  ({\lambda}^{\ell}_{u_1,u_2})^*$ 
is the adjoint of the  $\lambda$ obtained by exchanging its lower and upper indexes and taking the complex conjugate.
Together with the condition $\sum_{\ell_1, \ell_2} {\cal C}_{\ell_1, \ell_2} {\bar{\cal C}}^{\ell_1, \ell_2} = 1$, Eq.~\eqref{eq:isometry} automatically guarantees normalization of the HBT state at every level.
It has been shown in~\cite{Duan,BTvidal} that under these assumptions, tensor tree states allow for an efficient evaluation 
of observables and correlation functions.  In the generic case  each tensor can be chosen to be different from the others. Being 
interested in scale-invariant systems, it is natural to assume all the tensors $\lambda$  to be equal. 
In the rest of the paper we will follow the formalism described in~\cite{jova08}.

\subsection{Evaluation of local observables}

\paragraph{Single-site observables.} In the limit of large $n$, the physical properties of such states are
fully determined by the Completely Positive Trace preserving (CPT) channel  ${\cal S}$ of Fig.~\ref{fig:alberone}c).
It  transforms a single site density matrix of elements $[\rho]^{u}_{\ell}\equiv \langle \xi_\ell| \rho| \xi_u\rangle$ into a 2-sites states 
${\cal S}(\rho)$ of elements 
 \begin{eqnarray}\label{eq:fff}
  \langle \xi_{\ell_1}, \xi_{\ell_2} |   {\cal S}(\rho)| \xi_{u_1},\xi_{u_2}\rangle = 
  \sum_{k_1,k_2}  {\bar{\lambda}}^{u_1, u_2}_{k_1} \;  [\rho]^{k_1}_{k_2} \; {{\lambda}}^{k_2}_{\ell_1, \ell_2}\;.
 \end{eqnarray}  
Consider then a  family ${\cal F}\equiv \{ |\psi^{(n)}\rangle ; n =2, 3, \cdots \}$ of HBTSs of increasing sizes (depths)
sharing the same $\lambda$ and ${\cal C}$.
The map ${\cal S}$  
allows us to construct a  simple recursive expression for 
the reduced density operator \begin{eqnarray}
\bar{\rho}_1^{(n)} \equiv \frac{1}{N} \sum_{\alpha = 1}^{N} \rho^{(n)}_{\alpha}\;,
\end{eqnarray} 
which describes the averaged single site state 
of the $n$-th element of ${\cal F}$ (here $\rho^{(n)}_{\alpha}$ is the
reduced density matrix of the $\alpha$-th site of the system). 
Specifically  the isometric property of $\lambda$ yields,
\begin{equation} \label{eq:single}
 \bar{\rho}^{(n+1)}_1 = \Di ( \bar{\rho}_1^{(n)} )\;,
\end{equation}
where $\Di$ is the CPT map obtained by taking an equally weighted  mixture of  the partial traces of the
map ${\cal S}$ as indicated in Fig.~\ref{fig:alberone}b. This can be expressed as  
\begin{eqnarray} \Di  \equiv (\Di_L + \Di_R) / 2\;,
\end{eqnarray}
 where 
$\Di_L(\cdot)\equiv \mbox{Tr}_2[ \eS(\cdot) ]$ and $\Di_R(\cdot)\equiv \mbox{Tr}_1[ \eS(\cdot) ]$ with $\mbox{Tr}_{1,2}$ representing
partial traces with respect to the first and second site respectively. 
Equation~\eqref{eq:single} allows us  
to compute
the average expectation of a single site observable $\Theta$, for \emph{every} full depth value $n$ of the tree in terms of a repetitive application
of the map $\Di$. Indeed indicating with  $\langle \Theta_\alpha\rangle^{(n)}$  the expectation value on the $\alpha$-th site of $|\psi^{(n)}\rangle$
we can write
  \begin{eqnarray}
  \frac{1}{N}  \sum_{\alpha = 1}^{2^n} \langle  \Theta_{\alpha} \rangle^{(n)} 
 = \mbox{Tr}[  \Theta \; \bar{\rho}_1^{(n)} ]  =
  \mbox{Tr}[ \Theta \cdot \Di^{n-1} (\rho_{\text{hat}}) ]\;,
\end{eqnarray} 
where  $\rho_{\text{hat}} \equiv \bar{\rho}^{(1)}_1$ is  the single site density matrix of
elements  $\langle \xi_{\ell} | \rho_{\text{hat}}| \xi_{u} \rangle \equiv   \sum_{k} [ {\cal C}_{u,k}^* {\cal C}_{\ell,k}+  {\cal C}_{k,u}^* {\cal C}_{k,\ell}]/2$,
and where  $\Di^{n} \equiv \Di \circ \cdots \circ \Di$ with ``$\circ$" representing the composition of CPT maps.

\paragraph{Two-site observables.} An analogous procedure can be used to expressed averages of operators defined on $\nu=2$ neighboring sites.
All we have to do is to consider the density matrix
 \begin{eqnarray}
 \bar{\rho}^{(n)}_2 \equiv \frac{1}{N} \sum_{\alpha = 1}^{N} \rho_{\alpha, \alpha+1}^{(n)}\;,
 \end{eqnarray}
and build for this quantity a level-recursive mapping which is the 
two nearest-neighboring sites version of 
Eq.~\eqref{eq:single} (here  $\rho_{\alpha, \alpha+1}^{(n)}$ represents 
the reduced density matrix of the sites $\alpha$ and $\alpha+1$ 
associate with a HBTS  of depth $n$). The calculation
is straightforward so we just write the result,
\begin{equation} \label{eq:double}
 \bar{\rho}^{(n+1)}_2 = \frac{1}{2} (\Di_R \otimes \Di_L) (\bar{\rho}^{(n)}_2) +
 \frac{1}{2} \eS (\bar{\rho}_1^{(n)}).
\end{equation}
The above expression allows us to deal also with the case of observables operating on $\nu$ neighboring sites. Indeed for $\nu\geq 3$  it can be shown that any average density matrix 
$\bar{\rho}_\nu^{(n)}$ can be expressed in
terms of $\{ \bar{\rho}_2^{(m)}\}_{m < n}$ via the application of a proper CPT map deriving from $\eS$.
As an example, we write explicitely the case for $\nu = 3$ and $4$:
\begin{eqnarray} \label{eq:duetrequat}
 \bar{\rho}_3^{(n)} &=& {\textstyle \frac{1}{2}} \left[ \Di_R \otimes \eS + \eS \otimes \Di_L \right]
 (\bar{\rho}_2^{(n-1)})  \\
 \bar{\rho}_4^{(n)} &=& {\textstyle \frac{1}{2}} \left[ \eS \otimes \eS \right] (\bar{\rho}_2^{(n-1)}) +
 {\textstyle \frac{1}{4}} [\Di_R \otimes \eS \otimes \Di_L \nonumber] \circ \\
 && \circ \left[ \Di_R \otimes \eS + \eS \otimes \Di_L \right] (\bar{\rho}_2^{(n-2)}) \;.
\end{eqnarray}


\paragraph{Thermodynamic limit.} 
In the  limit of infinitely many sites, from Eq.~\eqref{eq:single} it follows that if
the average single site state associated with a HBTS of infinite depth characterized by a given isometry $\lambda$ is defined,
then it  must be a fixed point of the map $\Di$.
Since CPT maps have a unique fixed point except for a subset of  zero probability~\cite{MIX}, the fixed point is defined 
amost-always.
Similarly we can also provide an explicit formula for the thermodynamic limit of the 2-sites state~\eqref{eq:double}, i.e. 
\begin{eqnarray}
\bar{\rho}_2^{(\infty)} \equiv \lim_{n \to \infty} \bar{\rho}_2^{(n)}\;.
\end{eqnarray} This can be written either
as a self-consistent equation or like a series in terms of $\bar{\rho}_1^{(\infty)}$ by exploiting the identity~\eqref{eq:double}
 \begin{eqnarray}
 \bar{\rho}_2^{(\infty)} = \frac{1}{2} \sum_{m = 0}^{(\infty)} \left[ \frac{1}{2^{m}}
 \left( \Di_R \otimes \Di_L \right)^m \right] \circ \eS(\bar{\rho}_1^{(\infty)})\;.
 \end{eqnarray} 
The series in convergent in  any norm, thanks to the geometric factor
and the fact that CPT are non expansive. 
Such argument becomes even simpler when dealing with three or more n-n sites density matrices.
Indeed 
one can show that
 for any integer $\nu$
there  exists a CPT map $\Di_{2 \to \nu}$ such that, the
$\nu$ nearest neighbors sites density matrix $\bar{\rho}_{\nu}^{(\infty)}$ (averaged over translations) in the thermodynamic limit
is given by,
\begin{eqnarray}
\bar{\rho}_{\nu}^{(\infty)} = \Di_{2 \to \nu} (\bar{\rho}_2^{(\infty)})\;.
\end{eqnarray} 
This provides a complete characterization of the local properties of our
infinitely deep HBTS. For future reference we report 
the expression for case $\nu=3$ and $4$, 
\begin{eqnarray} 
  \Di_{2 \to 3} &=& \left( \Di_R \otimes \eS + \eS \otimes \Di_L \right)/2 \;,  \\
  \Di_{2 \to 4} &=& \left( \eS \otimes \eS + (\Di_R \otimes \eS \otimes \Di_L) \circ \Di_{2 \to 3} \right)/2\;,\label{eq:richiamino}
\end{eqnarray}
(notice that $\Di_{2\to 3}$ is exactly the channel which enters in Eq.~(\ref{eq:duetrequat})).
Finally we notice that all these quantities are independent from the element  ${\cal C}$ of the HBTS, implying that in the thermodynamical limit, the 
local structure of the state loses all its dependence from such element. As the physics of the system is determined by the  algebra of the local observables,  
this implies that {\em all} HBTS  of infinite depth, associated with a given $\lambda$ but with different ${\cal C}$ describe the {\em same} state.

\subsection{Correlation functions}
\label{sec:corr}  
Similarly to what has been done for the local observables in the previous section, also correlation function can be expressed 
in terms of iterative application of certain maps. Most important for our work is to show that this procedure leads naturally, in the 
case of homogeneous trees, to power-law decays for the correlators: the exponents being related to the eigenvalues of the 
map~\cite{jova08}.
 
In this following we  focus on  two-point correlation functions. 
As discussed before, since HBTSs are  not manifestly translationally invariant,
an average over translations has to be made. We thus introduce the quantities
\begin{eqnarray}
 \Cor^{(n)}_{\Delta\alpha} \equiv \frac{1}{2^n} \sum_{\beta = 1}^{2^n}
  [ \langle \Theta_{\beta} \, \Theta'_{\beta + \Delta\alpha} \rangle^{(n)} -
  \langle \Theta_{\beta} \rangle^{(n)} \langle \Theta'_{\beta + \Delta\alpha}
  \vphantom{\sum} \rangle^{(n)} ]\;, \nonumber 
  \end{eqnarray}
with $\Theta$ and $\Theta'$ being two single sites observables.
A remarkable simplification is achieved for any distance $\Delta\alpha$
equal to a power of 2. Under this condition we find that
 \begin{eqnarray} \Cor^{(n)}_{\Delta \alpha =2^m} =
 \mbox{Tr} [ (\Theta \otimes \Theta') \; \Dsla^{m} (
 \bar{\rho}_2^{(n-m)} - \bar{\eta}_{1,1}^{(n-m)}) ]\;,
 \end{eqnarray} 
where  $\Dsla$ is the map 
 $\Dsla \equiv \frac{1}{2} \left(
 \Di_L \otimes \Di_L + \Di_R \otimes \Di_R \right)$. The quantity 
 $\bar{\eta}_{1,1}^{(n)}$ is the averaged 2 sites nearest neighbour density matrix after
we traced away every quantum correlation, while keeping eventual
classical correlations intact, i.e.
 \begin{eqnarray}
 \bar{\eta}_{1,1}^{(n)} = \frac{1}{2^n} \sum_{\alpha = 1}^{2^n}
 \rho^{(n)}_{\alpha} \otimes \rho^{(n)}_{\alpha+1}\;.
 \end{eqnarray} 
Take then $n \to \infty$ while keeping $m = \log_2 \Delta\alpha$ fixed.
In this context it is important to notice that, like $\bar{\rho}_2^{(n)}$ also
$\bar{\eta}^{(n)}_{1,1}$ has a well-defined limit. It coincides with the two sites state,
 \begin{eqnarray}
 \bar{\eta}_{1,1}^{(\infty)}= \frac{1}{2} \sum_{m = 0}^{(\infty)} \left[ \frac{1}{2^{m}}
 \left( \Di_R \otimes \Di_L \right)^m \right] \circ (\Di_L \otimes \Di_R)(\sigsla)\;,
 \end{eqnarray} 
with $\sigsla$ being the fixed point of $\Dsla$. 
Exploiting this fact we can thus write the thermodynamic limit of the correlation  as 
\begin{eqnarray}
\Cor^{(\infty)}_{\Delta\alpha = 2^m} &=&
 \mbox{Tr} [ (\Theta \otimes \Theta') \;\; \Dsla^{m} ( \bar{\rho}_2^{(\infty)} - \bar{\eta}_{1,1}^{(\infty)})]
 \nonumber \\
 &=&
 \mbox{Tr} [( \bar{\rho}_2^{(\infty)} - \bar{\eta}_{1,1}^{(\infty)}) \;\; \Asla^{m}(\Theta \otimes \Theta')]\;,
 \end{eqnarray} 
where $\Asla$ is the adjoint superoperator of $\Dsla$ (with respect to the operator scalar product
$\langle A, B \rangle = \mbox{Tr}[A^{\dagger} B])$.
Recall that we are keeping track of the quantum correlations only,
moreover, since $\bar{\rho}_2^{(\infty)} - \bar{\eta}_{1,1}^{(\infty)}$
is a traceless matrix, we are guaranteed that
$\Cor^{(\infty)}_{\Delta\alpha} \to 0$ for ${\Delta\alpha} \to \infty$, because $\Dsla^{\infty} (X) = \sigsla \;\mbox{Tr}[X]$.
This shows that the only
residual influence on $m$ is kept through the number of times
we have to apply the appropriate map to its matrix argument.
By decomposing  $\Asla$ in Jordan blocks, one finds its set of eigenoperators;
let us assume that $\Theta \otimes \Theta'$ is one of such operators, related to the eigenvalue $\kappa$,
then the correlation function is expressed as follows
\begin{eqnarray} \Cor^{(\infty)}_{2^m} = g \; {\Delta\alpha}^{\log_2 \kappa }\;,
\end{eqnarray} 
 where we separated the prefactor
\begin{eqnarray} g = \Cor^{(\infty)}_{1} =
 \mbox{Tr} [( \bar{\rho}_2^{(\infty)} - \bar{\eta}_{1,1}^{(\infty)}) \;\; \Theta \otimes \Theta']\;.\end{eqnarray} 
The critical exponents are defined by the
spectrum of $\Asla$, and the related primary fields are given by the respective eigenoperators.
Notice that such exponents have always negative real-part, since all $|\kappa| \leq 1$
because $\Asla$ is a completely positive unital operator (and with the mixing condition only
$|\kappa| < 1$ and $\kappa = 1$ are allowed, e.g. see Ref.~\cite{NJP}); this guarantees that such correlations are
actually decaying power-law functons.

If the observables $\Theta \otimes \Theta'$ are not an eigenoperator of $\Asla$
their correlator is typically a sum of power-laws and may manifest logarythmic corrections
(arising from the fact that $\Asla$ is generally not diagonalizable)
\begin{eqnarray}
| \Cor^{(\infty)}_{\Delta\alpha} | \simeq \sum_{\kappa} \, {\Delta\alpha}^{\log_2 |\kappa| }
\;g_{\kappa}(\log_2 \Delta\alpha)\;,\end{eqnarray}
 where the sum spans over the spectrum of $\Asla$, and
$g_k (\cdot)$ are polynomials in their main argument.
The present considerations prove the criticality character of HBTS.

\section{Parent Hamiltonians}\label{sec:par} 
In the previous section we showed that HBTSs support power-law decay of correlators and we related the 
associated critical exponents to the tensors which the define the state. Is there any case where a HBTS is 
the {\em exact} ground state of a short-range critical Hamiltonian? In this section we show
how to construct local translationally invariant Hamiltonians for which
a given homogeneous HBTS is the exact  ground state. First we focuses on 
the case of infinite dimensional systems (thermodynamical limit). Then we show how the analysis can be
extended to the case of finite dimensional HBTSs.

\subsection{Thermodynamic limit}\label{sec:thermo} 

Consider a Hamiltonian which involves at most $(\nu-1)$-neighboring couplings of the form
\begin{eqnarray}
\mathcal{H} = \frac{1}{N} \sum_{\alpha = 1}^N H_{\nu}(\alpha)\;. \end{eqnarray} 
The factor $1/N$ is introduced  to keep a finite spectrum
in the thermodynamical limit,
and where $H_{\nu}(\alpha)$ is an interaction term that couples $\nu$ consecutive  sites starting from the  
$\alpha$-th (i.e.  the sites $\alpha, \cdots, \nu-1+ \alpha$). 
The expectation values
over the infinite HBTS of this Hamiltonian is
 \begin{eqnarray}
 \langle \mathcal{H} \rangle^{(\infty)} = \mbox{Tr} [ H_\nu \;  \bar{\rho}_{\nu}^{(\infty)} ]\;,\end{eqnarray} 
with $\bar{\rho}_{\nu}^{(\infty)}$
(the averaged $\nu$-neighboring sites density matrix)
being a quantity we can calculate  as discussed in the 
 previous sections. 

Let us for a moment assume that the rank of $\bar{\rho}_{\nu}^{(\infty)}$ is less than its
maximum $d^{\nu}$. This means that such density matrix has a nontrivial kernel,
which can be decomposed in a basis of vectors $\{ |\phi_{\nu}{(k)} \rangle \}_k$.
Therefore we take
 \begin{eqnarray}
 H_\nu= \sum_{k} E_k |\phi_\nu(k) \rangle \langle \phi_\nu(k) |\;, \label{ddd}
 \end{eqnarray}
 with $E_k$ being arbitrary positive constants. 
This is positive by construction, and so is the associated $\mathcal{H}$. Then, since the image
of $H_\nu$ belongs to the kernel of $\bar{\rho}_{\nu}^{(\infty)}$, it is clear that $H_\nu \;\bar{\rho}_{\nu}^{(\infty)} = 0$,
and so $\langle \mathcal{H} \rangle^{(\infty)} = 0$ as well. In the end, we built a positive, translation invariant, 
Hamiltonian, with $(\nu-1)$-neighboring coupling terms,  whose expectation value over our HBTS  is zero; this means
that the state is a ground state for $\mathcal{H}$.
The only caveat to make it work is to demonstrate that, for 
 some $\nu$ we have 
 \begin{eqnarray}\mbox{rank}(\bar{\rho}_{\nu}^{(\infty)}) < d^{\nu}\;, \end{eqnarray}
(otherwise $H_\nu$ would be the trivial null operator).
The fundamental ingredient to verify this  is to notice that the channel $\eS$ of Eq.~(\ref{eq:fff})  preserves rank
while increasing dimensions (i.e. it is an isometric mapping).
Let thus  investigate the case $\nu=3$. We know that the state $\bar{\rho}^{(\infty)}_3$  
  is obtained by exploiting the first of the mapping of Eq.~\eqref{eq:richiamino}. Specifically we have
 \begin{eqnarray}
 \bar{\rho}^{(\infty)}_3 = \Di_{2 \to 3} (\bar{\rho}_2^{(\infty)}) = ( {\cal I} \otimes \eS ) (A) + (\eS \otimes {\cal I} )(B)\;,\end{eqnarray} 
with ${\cal I}$ being the single site identity mapping and with  $A$ and $B$ some $d^2 \times d^2$ positive matrices. The maps ${\cal I} \otimes \eS$ and
$\eS \otimes {\cal I}$ preserve the rank,  and the rank of the sum is less or equal than
the sum of ranks,  thus leading us to the inequality
\begin{eqnarray}
 \mbox{rank}(\bar{\rho}_3^{(\infty)}
 ) \;\leq\; 2\,d^2\;,\end{eqnarray}
over a maximum of $d^3$. Therefore  if the local dimension $d$ is 3 (spin 1) or greater
then we already achieved our goal of finding a $\bar{\rho}_{\nu}^{(\infty)}$ matrix with non-maximal
rank. 

For  $d=2$ (spin 1/2) instead  we have to 
move to $\nu = 4$.
In this case the state to consider is 
\begin{eqnarray}
\bar{\rho}^{(\infty)}_4 & = & \Di_{2 \to 4} (\bar{\rho}_2^{(\infty)})\nonumber\\
& = &(\eS \otimes \eS)  (A') + ( {\cal I} \otimes \eS \otimes {\cal I}) (B')\;.
\end{eqnarray} 
Since its rank  obeys  the inequality 
\begin{eqnarray}
\mbox{rank}(\bar{\rho}^{(\infty)}_4) \leq d^2 + d^3\;,\end{eqnarray}   we have found a state  that 
already for $d = 2$ possess a nontrivial kernel (indeed in this case  $\mbox{rank}(\bar{\rho}^{(\infty)}_4) \leq 12$
which is strictly minor than the total dimension $d^4 = 16$).

In summary this shows that any given infinite HBTS admits always a local translationally invariant non-trivial
parent Hamiltonian ${\cal H}$, which can be constructed explicitly as in Eq.~(\ref{ddd}).
For $d\geq 3$ such ${\cal H}$ can be chosen to have interactions which involve up to
second neighboring couplings. For $d=2$ instead we can always choose ${\cal H}$ with up to
third neighboring couplings.
Our analysis deals with the worst case scenario,
if it occurs that $\bar{\rho}^{(\infty)}_2$ is non full rank by accident,
one can construct a shorter-ranged (i.e. nearest neoughbours) parent hamiltonian using ~(\ref{ddd}).

\subsection{Finite size case} \label{sec:finite}
The above approach formally applies to the case of infinitely many  sites,  and in general there is  no
guarantee that the selected ${\cal H}$ will be  a parent Hamiltonian of the HBTS $|\psi^{(n)}\rangle$ when $n$ is finite.
Nonetheless the proof can be extend to cover also this case in most situations.
This will allow us to prove that for $N$ even, ${\cal H}$ is unfrustrated and that its ground space
must have  dimension $D_{\text{gr}}$ larger than $d^{N/2}$.

To show this we focus on the case  $d\geq 3$ and assume  that our HBTS
has $\bar{\rho}_2^{(\infty)}$ of full rank (generalization to $d=2$ shall be dealt later);
this guarantees that Eq. ~(\ref{ddd}) provides a parent hamiltonian  ${\cal H}$ for the thermodynamical state
with a three-body interaction $H_3$.
Consider then a generic state $|\psi\rangle$  of $N/2$ sites and ``grow''  a BT level from it, using the same $\lambda$ isometry
we used to build  ${\cal H}$. This way we obtain a $N$-sited state
\begin{eqnarray} \label{statophi}
|\phi\rangle = \lambda^{\otimes N/2} |\psi\rangle\;,
\end{eqnarray}  which, by varying $|\psi\rangle$, spans a subspace $\gspa$ of dimension  $d^{N/2}$
(when $N$ is power of 2 one element of such subspace is for instance the HBTS we started with).  
The expectation value $\langle \phi | {\cal H} | \phi \rangle$ can then be expressed as 
\begin{eqnarray} \label{eq:growthy}
  \mbox{Tr} [ \bar{q}_3
  \; H_3 ] =
 \mbox{Tr}[ \Di_{2 \to 3} (\bar{r}_2 ) \; H_3 ] 
 \nonumber   =
 \mbox{Tr} [ \bar{r}_2  \; \Ai_{3 \to 2}( H_3 ) ]\;,
\end{eqnarray}
where $\bar{q}_3$ is the averaged reduced density matrices of  3-neighboring sites of $|\phi\rangle$,
$\bar{r}_2$ is the averaged reduced density matrices of  2-neighboring sites of $|\psi\rangle$, while 
 $\Ai_{3 \to 2}$ is the Heisenberg conjugate map  of $\Di_{2 \to 3}$. 
At this point we observe that  $\Ai_{3 \to 2}\left( H_3 \right)$ is the null
operator. This follows form Eq.~(\ref{eq:richiamino}) which allows us to write
\begin{figure}
\includegraphics[width=\linewidth]{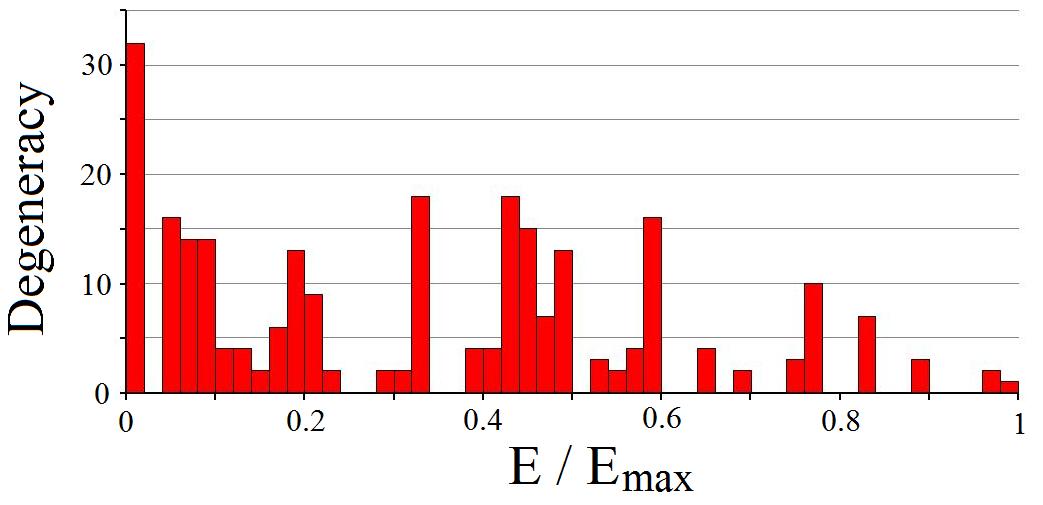} 
\caption{(Color online) Unnormalized density of states  of the parent Hamiltonian
generated from a sample HBTS for  $N = 8$ sites 
(energy levels have been rescaled to the maximum energy eigenvalue).
For this example it occurs that the ground state degeneracy is twice the lower bound we discuss
in the paper: 32-fold over 256 states. More precisely, we  have shown that the ground space of the Hamiltonian in this case coincides with  the direct sum $\gspa \oplus T(\gspa)$ of 
the space $\gspa$ formed by the vectors defined in Eq.~(\ref{statophi}), and by its translated version $T(\gspa)$, $T$ being the one-site translation
(we verified that in this case the  two sectors form linearly independent subspace).}
\label{fig:spettro}
\end{figure}
\begin{eqnarray}
 0 = \mbox{Tr} [ \bar{\rho}_3^{(\infty)} \;  H_3 ]  &= 
 \mbox{Tr} [ \Di_{2 \to 3}(\bar{\rho}_2^{(\infty)}) \; H_3 ] \nonumber \\
  &=  \label{proofv}
 \mbox{Tr} [ \bar{\rho}_2^{(\infty)} \Ai_{3 \to 2}(H_3) ],
\end{eqnarray}
where the first identity simply states that ${\cal H}$ is the parent Hamiltonian of the HBTS at thermodynamical limit.
Since   $\bar{\rho}_2^{(\infty)}$  has maximal support by 
hypothesis and $\Ai_{3 \to 2}(H_3)$ is positive semidefinite  by construction, Eq.~(\ref{proofv}) implies 
$\Ai_{3 \to 2}(H_3)=0$. 
 Equation~\eqref{eq:growthy}  then leads to
$ \langle \phi | {\cal H} | \phi \rangle =0$  which, together with the
positivity of ${\cal H}$, tells us that each one of the vectors $|\phi\rangle$ of the subspace $\gspa$ 
is a ground state of the parent Hamiltonian ${\cal H}$.

Let us now deal briefly with the case $d=2$. If $\bar{\rho}_3^{(\infty)}$ is nonfull rank
then we can build a three-body interacting parent hamiltonian just like the $d \geq 3$ case,
and the generalization to the finite setting is identical. Otherwise,  via $\bar{\rho}_4^{(\infty)}$
we can build a positive parent hamiltonian  ${\cal H}$ 
of the thermodynamical state with four-body
interactions $H_4$. Evaluating its expectation value on the  $N$-sited state~(\ref{statophi})
  it is then easy to verify that 
it nullify (the proof is similar to the previous case, and it exploits the fact
that $\bar{\rho}_3^{(\infty)}$ has full rank). 

The above discussion proves that for all even  $N$, the Hamiltonian 
${\cal H}$ (resp. ${\cal H}'$) has a ground eigen-space which is at least $d^{N/2}$ dimensional~\cite{COMMENT}.
The presence of a wide ground state degeneracy is in accordance with symmetry predictions: since
a finite HBT state $|\psi\rangle$ breaks the translational symmetry at every lenghtscale, the
whole space generated by $\left\{ |\psi\rangle, T |\psi\rangle, T^2|\psi\rangle \ldots T^N|\psi\rangle \right\}$
must be embedded within the ground space.
The present argument also implies  that ${\cal H}$ represents an unfrustrated system. Indeed if
  $ \langle \phi| {\cal H} | \phi \rangle =0$  then
  each local component  of ${\cal H}$ needs to nullify  on $|\phi\rangle$,  i.e.  $\langle \phi | H_{\nu}(\alpha) | \phi \rangle =0$ $\forall \alpha$. 
As an example in Fig.~\ref{fig:spettro} we report the eigenvalues  degeneracies  for 
 a parent Hamiltonian ${\cal H}$ generated from an isometry $\lambda$ defined   
by the following mapping 
 \begin{eqnarray}
 |0\rangle &\rightarrow&  |01\rangle\;, \\ 
 |1\rangle &\rightarrow&
 \frac{1}{\sqrt{2}} ( |00\rangle + |11\rangle)\;,   \end{eqnarray}
  (here $d=2$ while ${\cal H}$ was generated by taking the free-parameters $E_k$ of Eq.~(\ref{ddd}) to be uniform). 
For $N=4,6,8$ the ground state degeneracy turns out to be exactly  $2 \; 2^{N/2}$ showing that in this case $\gspa$ and 
$T(\gspa)$~\cite{COMMENT}  saturate completely the corresponding eigenspace   (the figure only reports the case $N=8$).  
We also checked numerically the case of $N$ odd (for which the previous theoretical analysis does not hold):  in this case the ground state energy is not null showing that ${\cal H}$ is frustrated and that its degeneracy is way smaller than $d^{N/2}$,
and controllable by choosing the $E_k$ parameters appropriately.

This procedure to construct a parent Hamiltonian for a binary tree can also be applied to other tensor 
structures.  In the Appendix we discuss the case of a MERA.

\section{Conclusions}
\label{con}
In this paper we analyzed the potential  of binary trees to simulate efficiently quantum critical systems.
Previous works~\cite{BTvidal} presented evidences that such states  could yield violation of  the area law with logarithmic corrections. 
In this paper we focused on homogenous configurations which
allow for an explicit analytic treatment of the thermodynamical limit. 
Their hierarchical,  scale invariant structure suggest that  they should be capable of 
exhibit critical behaviors, at least once a proper  averaging over translations has been performed to compensate for their explicit lack 
of translational invariance~\cite{NOTA1}.
For instance by looking at their tensorial decomposition it is clear that HBTS
do not violate the area law for all possible partitions of the sites (e.g. since the left side of the graph is connected with the right side by only a
single link, the resulting block entropy will be independent from the number of sites). It is reasonable to belief that such
 "anomalies" however will wash away when averaging  over all possible translations (a legitimate operation when  simulating translationally invariant systems). 
In the case of the block entropy this can be heuristically verified by noticing that indeed the average number of tensor links that needs to be
cut in order to disconnect the causal cone of a block of consecutive sites from the rest,  scales almost logarithmically with the block size.
To  test the validity of these arguments, in  our paper we focused on the behavior of two-point correlation functions in the thermodynamical
limit of infinitely many sites. Once averaged over all possible translations, we proven that these quantities can be explicitly  computed and 
showed that they decay as power law in agreement with the criticality character of HBTS. 

In the second part of the paper we have then shown that HBTS are the exact ground states of short-range interacting Hamiltonians. In particular we gave a  procedure to built  such parent Hamiltonian. In the case we analyzed the ground state has a degeneracy 
which scales with the square root of the dimension of the Hilbert space.  Similar results can be obtained (see the appendix) also 
for MERA states.

\acknowledgments
We acknowledge fruitful discussions with G. E. Santoro and financial support from IP-EUROSQIP, FIRB-RBID08B3FM, 
SFB/TRR 21, and the National Research Foundation and Ministry of Education Singapore.

\appendix

\section{Parent Hamiltonian for MERA states} 
\label{meraparent}
In this Appendix we discuss how to generalize the analysis of Sec.~\ref{sec:par}  to the case of scale-invariant (i.e. homogeneous)
MERA states ~\cite{mera}. Indeed also for a MERA it is possible to  establish upper bounds for the rank of the states
$\bar{\rho}_{\nu}^{(\infty)}$ (the translationally averaged $\nu$ neighbouring sites density matrix in
the thermodynamical limit) by exploiting growth superoperators properties, for varous block sizes $\nu$.
Therefore, by finding the suitable (smallest) $\nu$ for which such rank is not maximal, the construction of a parent
hamiltonian interaction term according to equation \eqref{ddd} is straightforward.\\
Precisely, such minimal parent interaction lenght $\nu$ depends on the topology of the original MERA
~\cite{Mera23} and its local dimension $d$. For a binary MERA structure, we find that
$\text{rank}(\bar{\rho}_{5}^{(\infty)}) \leq 2 d^4$
whose value is not maximal for $d \geq 3$; for completeness
$\text{rank}(\bar{\rho}_{6}^{(\infty)}) \leq d^4 + d^5$
takes care of the case $d = 2$; thus is always possible to build a parent hamiltonian
with a 5 or 6 bodies interaction.
When considering to a ternary MERA structure, we have to involve a block of seven nearest neighbours to achieve
the meaningful bound
$\text{rank}(\bar{\rho}_{7}^{(\infty)}) \leq 3d^5$, always nonmaximal regardless of $d$.
Furthermore, one can produce analogous conditions under which
${\cal H}$ will be still a parent (unfrustrated) Hamiltonian for the finite site case and
verify that it posses a ground state degeneracy which is exponentially large (order $d^{N/2}$ or $d^{N/3}$).
A main difference between this case and the previous one is that a parent hamitonian for a finite HBT
state also admits always a dimerized ground state (just pick up a vector of Eq.~(\ref{statophi}) that is build by taking $|\psi\rangle$ as a product state),
while in the MERA context there is no such proof of triviality.


\begin{thebibliography}{1}
\bibitem{COMMENT} 
 Clearly, our argument does not necessarily provide a complete characterization of the
 ground space, as $D_{\text{gr}}$ can be even larger of $d^{N/2}$. In particular it is possible that the space $\gspa$ will not be
 invariant under translation by one site (notice though, that if $T$ is the translation by one site, 
 the subspace $\gspa$ is explicitly invariant under $T^2$).
\bibitem{NOTA1} Notice that such averaging needs not to be performed in a classical fashion (i.e. by replacing the state $|\psi^{(n)}\rangle$ 
with a mixture of all possible translations of the same vector). Instead the same average can be performed in a coherent fashion
by replacing $|\psi^{(n)}\rangle$  with a pure state $|\bar{\psi}^{(n)}\rangle$ obtained by coherent superimposing all its translated versions. 
On the base of "ultraviolette catastrophe" arguments, we can  indeed argue that, in the thermodynamical limit, the statistics associated with any local
observable will be identical in the two cases.
\end{thebibliography}
\end{document}